\documentclass[journal]{IEEEtran}
\usepackage{amsmath}
\usepackage{graphicx}
\usepackage{caption2}
\usepackage{amsthm}
\usepackage{setspace}
\usepackage{mathtools}
\begin{document}
\title{A New Proof for the DoF Region of the MIMO Networks with No CSIT}
%\author{\IEEEauthorblockN{Borzoo Rassouli\IEEEauthorrefmark{1}, Chenxi Hao\IEEEauthorrefmark{1} and Bruno Clerckx\IEEEauthorrefmark{1}\IEEEauthorrefmark{2}}\\
%\IEEEauthorblockA{\IEEEauthorrefmark{1}EEE%\\Imperial College London, United Kingdom
%}\\
%\IEEEauthorblockA{\IEEEauthorrefmark{2}EEE %Korea University, Korea\\
%\\
%Email: \{Author 1; Author 2; Author 3\}@}
%%Email: \{b.rassouli12; chenxi.hao10; b.clerckx\}@imperial.ac.uk}
%}
%\maketitle
\author{{Borzoo Rassouli, Chenxi Hao and Bruno Clerckx} %
\thanks{Borzoo Rassouli, Chenxi Hao and Bruno Clerckx are with the Communication and Signal Processing group of Department of Electrical and Electronics,
Imperial College London, United Kingdom. emails: \{b.rassouli12; chenxi.hao10; b.clerckx\}@imperial.ac.uk}
\thanks{Bruno Clerckx is also with the School of Electrical Engineering, Korea University, Korea.}
\thanks{This work was partially supported by the Seventh Framework Programme for Research of the European Commission under grant number HARP-318489.}}
\maketitle
\begin{abstract}
%\boldmath
In this paper, a new proof for the degrees of freedom (DoF) region of the $K$-user multiple-input multiple-output (MIMO) broadcast channel (BC) with no channel state information at the transmitter (CSIT) and perfect channel state information at the receivers (CSIR)  is provided. Based on this proof, the capacity region of a certain class of MIMO BC with channel distribution information at the transmitter (CDIT) and perfect CSIR is derived. Finally, an outer bound for the DoF region of the $K$-user MIMO interference channel (IC) with no CSIT is provided.
\end{abstract}

\newtheorem{theorem}{Theorem}
\section{Introduction}
Spatial multiplexing is a key feature of MIMO communication networks \cite{Bruno}. The DoF region, which is the capacity region normalized by the logarithm of SNR in high SNR regimes, is a metric that captures the spatial multiplexing property. The DoF region of the MIMO BC with no CSIT was first shown in \cite{Huang0}, \cite{Huang1} for the two user case and later in \cite{Vaze} for the general $K$-user BC.

In this paper, we provide an alternative proof for the results obtained in the mentioned papers based on a simple lemma where its advantage over \cite{Vaze} is in extending the results of \cite{Huang1} for the capacity region of special two-user broadcast channels to special K-user BCs. The paper is organized as follows. Section \ref{system} introduces the system model. The alternative proof is provided in section \ref{proof}. Subsequently, The capacity region of a certain $K$-user MIMO BC with CDIT and an outer bound for the DoF region of the MIMO IC with no CSIT are provided in section \ref{capacity} and section \ref{IC}, respectively.

Throughout the paper, $f\sim o(\log P)$ is equivalent to $\lim_{P\to\infty}\frac{f}{\log P}=0$ and for a pair of integers $m\leq q$, the discrete interval is defined as $[m:q]=\{m,m+1,\ldots,q\}$. $Y_{[i:j]}=\{Y_i,Y_{i+1},\ldots,Y_j\}$, $Y([i:j])=\{Y(i),Y(i+1),\ldots,Y(j)\}$ and $Y^n=Y([1:n])$.
$R_{\geq 0}$ denotes the set of non-negative real numbers.
\section{System Model and Main Results}\label{system}
We consider a MIMO BC, in which a transmitter with $M$ antennas sends independent messages $W_1,\ldots,W_K$ to $K$ users (receivers), where each receiver is equipped with $N_i$ receive antennas ($i\in[1:K]$). In a flat fading scenario, the discrete-time baseband received signal of user $i$ at channel use (henceforth, time slot) $t$ can be written as
\begin{equation}\label{e1}
  \tilde{\textbf{{Y}}}_i(t)=\textbf{{H}}_i^H(t)\textbf{{X}}(t)+\textbf{{Z}}_i(t)\ ,\ i\in[1:K]\ ,\ t\in[1:n]
\end{equation}
where $\textbf{{X}}(t)\in C^{M\times 1}$ is the transmitted signal satisfying the (per codeword) power constraint $\sum_{t=1}^{n}\|\mathbf{x}(t)\|^2\leq nP$. $\textbf{{H}}_i(t)\in C^{M\times N_i}$ and $\textbf{{Z}}_i(t)\in C^{N_i\times 1}$ are, respectively, the channel matrix and the additive noise vector of receiver $i$. The elements of $\textbf{{H}}_i(t)$ are independent identically distributed across time and users. The noise vectors and the elements of the channel matrices are allowed to have any tempered distribution (independent of $\textbf{X}(t)$) and the channel matrices are assumed to be full rank almost surely. We assume no channel state information at the transmitter and perfect local channel state information at the receiver (CSIR)  i.e., at time slot $t$, user $i$ has perfect knowledge of $\textbf{{H}}_i([1:t])$.

The rate tuple $(R_1,R_2,\ldots,R_K)$, in which $R_i = \frac{\log (|W_i|)}{n}$, is achievable if there exists a coding scheme such that the probability of error in decoding $W_i$ at user $i (i\in[1:K])$ can be made arbitrarily small with sufficiently large coding block length. Analysis of the capacity region $C(P)$, which is the closure of the set of achievable rate tuples, is not always tractable. Instead, we consider the DoF region, which is a simpler metric independent of the transmit power, and is defined as $\{(d_1,\ldots,d_K)|\exists (R_1,R_2,\ldots,R_K)\in C(P) \mbox{\ such that\ } d_i=\lim_{P\to \infty}\frac{R_i}{\log P} \forall i\}$. At very high SNRs, the effect of additive
noise can be neglected and what remains is the interference caused by other users' signals. Therefore, the DoF region could also be interpreted as the region constructed by the number of interference-free private data streams that users receive simultaneously per channel use.

\begin{theorem}
The DoF region of the $K$-user MIMO BC with no CSIT and perfect CSIR is given by
\begin{equation}\label{e2}
 D=\{(d_1,d_2,\ldots,d_K)\in R_{\geq 0}^K| \sum_{i=1}^K\frac{d_i}{r_i}\leq 1\}
\end{equation}
where $r_i=\min\{M,N_i\}$.
\end{theorem}
\section{Proof of the theorem 1}\label{proof}
Unlike \cite{Huang0} and \cite{Huang1}, the proof is not based on the degradedness of the MIMO BC under no CSIT. Without loss of generality, we assume $N_1\geq N_2\geq\ldots\geq N_K$ and we enhance the channel by giving the messages of users $[i+1:K]$ to user $i$. We also assume that each user not only knows its own channel, but also has perfect knowledge of the other users' channels. In other words, perfect global CSIR is assumed. It is obvious that this assumption does not reduce the outer bound which means that the bound with local CSIR is inside the bound with global CSIR; however, the achievability is based on only local CSIR. The region is further enhanced by giving all the noise vectors to each user. According to the Fano's inequality
\begin{equation}\label{e3}
  nR_i \leq I(W_i;\tilde{\textbf{Y}}_i^n|\Omega^n,\Lambda^n,W_{[i+1:K]})+n\epsilon_n\ \ ,\ i\in[1:K]
\end{equation}
where $W_{K+1}=\emptyset$ and $\epsilon_n$ goes to zero as $n$ goes to infinity. $\Omega^n$ is the global channel state information up to time slot $n$ and $\Lambda^n$ denotes the set of all the noise vectors across the users (extended over $n$ time slots). Let $S_i$ denote the index set of the $r_i(=\min\{M,N_i\})$ linearly independent elements of the $N_i$-dimensional vector $\textbf{{H}}_i^H\textbf{{X}}$ (note that $S_i$ is not necessarily unique). We decompose the $N_i$-dimensional received signal of user $i$ as $\tilde{\textbf{Y}}_i=(\textbf{Y}_i,\hat{\textbf{Y}}_i)$ where $\textbf{Y}_i$ corresponds to the set of $r_i$ linearly independent elements having their index in $S_i$ , i.e. $\textbf{Y}_i=\tilde{\textbf{Y}}_{i,S_i}$, and $\hat{\textbf{Y}}_i$ can be reconstructed by linear combination of the elements in $\textbf{Y}_i$ within noise level.
From the chain rule of mutual information,
\begin{align}\label{e4}
  nR_i &\leq I(W_i;\textbf{Y}_i^n|\Omega^n,\Lambda^n,W_{[i+1:K]})\nonumber \\
  &\ \ \ +\underbrace{I(W_i;\hat{\textbf{Y}}_i^n|\Omega^n,\Lambda^n,W_{[i+1:K]},\textbf{Y}_i^n)}_{o(\log P)}+n\epsilon_n.
\end{align}
For simplicity, we ignore $n\epsilon_n$ (since later it will be divided by $n$ and $n\to\infty$) and the term with $o(\log P)$ and write
\begin{align}
\sum_{i=1}^K\frac{nR_i}{r_i}  &\leq  \sum_{i=1}^K\frac{I(W_i;\textbf{Y}_i^n|\Omega^n,\Lambda^n,W_{[i+1:K]})}{r_i} \nonumber \\
 &\leq \underbrace{\frac{h(\textbf{Y}_K^n|\Omega^n,\Lambda^n)}{r_K}}_{\leq n\log P} +\sum_{i=1}^{K-1}\left[\frac{h(\textbf{Y}_i^n|\Omega^n,\Lambda^n,W_{[i+1:K]})}{r_i} \right.\nonumber\\
   &\left.\  \ \ -\frac{h(\textbf{Y}_{i+1}^n|\Omega^n,\Lambda^n,W_{[i+1:K]})}{r_{i+1}}\right] \label{e5}
\end{align}
where we have used the fact that $\frac{h(\textbf{Y}_1^n|\Omega^n,\Lambda^n,W_{[1:K]})}{r_1}\sim o(\log P)$, since with the knowledge of $\Omega^n,\Lambda^n,W_{[1:K]}$, the observation $\textbf{Y}_1^n$ can be reconstructed.
Before going further, the following lemma is needed. The authors in \cite{Shearer} prove the following lemma in a combinatorial theoretic approach, while our proof is based on induction and simple properties of entropy.\\
\textbf{Lemma}. Let $\Gamma_N=\{Y_1,Y_2,\ldots,Y_N\}$ be a set of $N(\geq2)$ arbitrary random variables and $\Psi_i^{j}(\Gamma_N)$ be a sliding window of size $j$ over $\Gamma_N$ ($1\leq i,j \leq N$) starting from $Y_i$ i.e.,
\[\Psi_i^{j}(\Gamma_N) = Y_{(i-1)_N+1},Y_{(i)_N+1},\ldots,Y_{(i+j-2)_N+1}\]
where $(.)_N$ defines the modulo $N$ operation. Then,
%\begin{multline}\label{e6}
 % (N-m)h(Y_1,Y_2,\ldots,Y_N|A)\leq \sum_{i=1}^{N}h(\Psi_i^{N-m}(\Gamma_N)|A)\nonumber\\
 %  1\leq m\leq N-1
%\end{multline}
\begin{equation}\label{e..6}
  (N-m)h(Y_{[1:N]}|A)\leq \sum_{i=1}^{N}h(\Psi_i^{N-m}(\Gamma_N)|A)
\end{equation}
where $m\in [1:N-1]$ and $A$ is an arbitrary random variable.
\begin{IEEEproof} We prove the lemma by showing that for every fixed $m(\geq 1)$, (\ref{e..6}) holds for all $N(\geq m+1)$ using induction. It is obvious that for every $m(\geq 1)$, (\ref{e..6}) holds for $N=m+1$. In other words, $h(Y_1,Y_2,\ldots,Y_N|A)\leq \sum_{i=1}^{N}h(Y_i|A)$. Now, considering that (\ref{e..6}) is valid for $N(\geq m+1)$, we show that it also holds for $N+1$. Replacing $N$ with $N+1$, we have
\begin{align}
  &(N+1-m)h(Y_{[1:N+1]}|A)\nonumber\\&=h(Y_{[1:N+1]}|A)+\!(N-m)h(Y_{[1:N-1]},\overbrace{Y_N,Y_{N+1}}^{Z}|A)\nonumber\\
&\leq h(Y_{[1:N+1]}|A)+\sum_{i=1}^{N}h(\Psi_i^{N-m}(\Phi_N)|A)\label{e..8}\\
  &= h(Y_{[1:N+1]}|A)+\sum_{i=1}^{m}h(\Psi_i^{N-m}(\Phi_N)|A)\nonumber\\
  &\ \ \ + \sum_{i=m+1}^{N}h(\Psi_i^{N+1-m}(\Gamma_{N+1})|A)\label{e..9}\\
  &= h(Y_{[N-m+1:N]}|Y_{N+1},Y_{[1:N-m]},A)\nonumber\\
&\ \ \ +\sum_{i=1}^{m}h(\Psi_i^{N-m}(\Phi_N)|A)+h(Y_{N+1},Y_{[1:N-m]}|A)\nonumber\\
&\ \ \ + \sum_{i=m+1}^{N}h(\Psi_i^{N+1-m}(\Gamma_{N+1})|A)\label{e..11}\\
&= h(Y_{[N-m+1:N]}|Y_{N+1},Y_{[1:N-m]},A)\nonumber\\
&\ \ \ +\sum_{i=1}^{m}h(\Psi_i^{N-m}(\Phi_N)|A)+\sum_{i=m+1}^{N+1}h(\Psi_i^{N+1-m}(\Gamma_{N+1})|A)\nonumber\\
&= \sum_{i=1}^mh(Y_{N-m+i}|Y_{N+1},Y_{[1:N-m+i-1]},A)\nonumber\\
&\ \ \ +\sum_{i=1}^{m}h(Y_{[i:N-m+i-1]}|A)+\sum_{i=m+1}^{N+1}h(\Psi_i^{N+1-m}(\Gamma_{N+1})|A)\label{e..13}\\
&\leq \sum_{i=1}^mh(Y_{N-m+i}|Y_{[i:N-m+i-1]},A)+\sum_{i=1}^{m}h(Y_{[i:N-m+i-1]}|A)\nonumber\\&\ \ \ +\sum_{i=m+1}^{N+1}h(\Psi_i^{N+1-m}(\Gamma_{N+1})|A)\label{e13..75}\\
&= \sum_{i=1}^{m}h(\Psi_i^{N+1-m}(\Gamma_{N+1})|A)+\sum_{i=m+1}^{N+1}h(\Psi_i^{N+1-m}(\Gamma_{N+1})|A) \nonumber\\
&= \sum_{i=1}^{N+1}h(\Psi_i^{N+1-m}(\Gamma_{N+1})|A)
\end{align}
%\begin{align}
%  &= \sum_{i=1}^{m}h(\Psi_i^{N+1-m}(\Gamma_{N+1})|A)+\sum_{i=m+1}^{N+1}h(\Psi_i^{N+1-m}(\Gamma_{N+1})|A) \label{e14}\\
%  &= \sum_{i=1}^{N+1}h(\Psi_i^{N+1-m}(\Gamma_{N+1})|A)
%\end{align}
where in (\ref{e..8}), $\Phi_N=\{Y_{[1:N-1]},Z\}$ and we have used the validity of (\ref{e..6}) for $N$. In (\ref{e..9}), we have used the fact that $\Psi_i^{N+1-m}(\Gamma_{N+1})=\Psi_i^{N-m}(\Phi_N)$ for $i \in [m+1:N]$ . In (\ref{e..11}), the chain rule of entropies is used and in (\ref{e..13}), the sliding window is written in terms of its elements. Finally, in (\ref{e13..75}), the fact that conditioning does not increase the differential entropy is used. Therefore, since $m(\geq 1)$ was chosen arbitrarily and (\ref{e..6}) is valid for $N=m+1$ and from its validity for $N(\geq m+1)$ we could show it also holds for $N+1$, we conclude that (\ref{e..6}) holds for all values of $m$ and $N$ satisfying $1\leq m\leq N-1$. \qedhere
\end{IEEEproof}
Each term in the summation of (\ref{e5}) can be written as
\begin{equation}\label{e15.75}
    \frac{r_{i+1}h(\textbf{Y}_i^n|S_{i,n})\!-\!r_ih(\textbf{Y}_{i+1}^n|S_{i,n})}{r_ir_{i+1}}
\end{equation}
\begin{align}
&= \frac{r_{i+1}h(\textbf{Y}_{i,1}^n,\textbf{Y}_{i,2}^n,\ldots,\textbf{Y}_{i,r_i}^n|S_{i,n})}{r_ir_{i+1}}-\frac{\!r_ih(\textbf{Y}_{i+1}^n|S_{i,n})}{r_ir_{i+1}}\\
&\leq \frac{\sum_{p=1}^{r_i}h(\Psi_p^{r_{i+1}}(\Gamma_{r_i})|S_{i,n})}{r_ir_{i+1}}-\frac{r_ih(\textbf{Y}_{i+1}^n|S_{i,n})}{r_ir_{i+1}} \label{e16}\\
&= \sum_{p=1}^{r_i}\left[\frac{h(\Psi_p^{r_{i+1}}(\Gamma_{r_i})|S_{i,n})}{r_ir_{i+1}}-\frac{h(\textbf{Y}_{i+1}^n|S_{i,n})}{r_ir_{i+1}} \right]\label{e17}\\
  &= \sum_{p=1}^{r_i}\left[\frac{h(\textbf{A}_{p,i,n}\textbf{X}^n+\textbf{B}_{p,i,n}|S_{i,n})}{r_ir_{i+1}}-\frac{h(\textbf{C}_{i,n}\textbf{X}^n+\textbf{D}_{i,n}|S_{i,n})}{r_ir_{i+1}} \right]\label{e17.5}\\
  &=0 \label{e18}
\end{align}
where $S_{i,n}=\{\Omega^n,\Lambda^n,W_{[i+1:K]}\}$ and in (\ref{e16}), since $r_{i+1}\leq r_i$, the result of the previous lemma is applied in which $\Gamma_{r_i}=\{\textbf{Y}_{i,1}^n,\textbf{Y}_{i,2}^n,\ldots,\textbf{Y}_{i,r_i}^n\}$ is the set of $r_i$ linearly independent elements in $\textbf{Y}_i^n$. In (\ref{e17.5}), we write $\Psi_p^{r_{i+1}}(\Gamma_{r_i})$ and $\textbf{Y}_{i+1}^n$ as large $nr_{i+1}$ dimensional vectors as follows. $\Psi_p^{r_{i+1}}(\Gamma_{r_i})=\textbf{A}_{p,i,n}\textbf{X}^n+\textbf{B}_{p,i,n}$ and $\textbf{Y}_{i+1}^n=\textbf{C}_{i,n}\textbf{X}^n+\textbf{D}_{i,n}$ where $\textbf{A}_{p,i,n}$ and $\textbf{C}_{i,n}$ ($\in C^{nr_{i+1}\times nM}$) capture the channel coefficients over the $n$ time slots, $\textbf{X}^n$ is the $nM$ dimensional input vector and $\textbf{B}_{p,i,n}$ and $\textbf{D}_{i,n}$ capture the noise vectors over the $n$ time slots. Since $\textbf{A}_{p,i,n}$ and $\textbf{C}_{i,n}$ are identically distributed channel coefficients and the noise terms are provided at each user, the arguments of the differential entropies in (\ref{e17.5}) are statistically equivalent (i.e., have the same probability density function) which results in (\ref{e18}). Therefore, (\ref{e5}) is simplified to
\begin{equation}
  \sum_{i=1}^K\frac{nR_i}{r_i}\leq n\log P.
\end{equation}
After dividing both sides by $n\log P$ and taking the limit $n,P \to \infty$, we get
\begin{equation}
  \sum_{i=1}^K\frac{d_i}{r_i}\leq 1.
\end{equation}
The above DoF region is achieved by a simple time sharing across the users where only local CSIR assumption is necessary.

Note that since the noise can be non-Gaussian, Gaussian codes may no longer be DoF-achieving. Finally, it can be observed that the assumption of independent channels across the users was not used in the proof and since it
does not change the achievability, it is not a necessary condition and can be relaxed.
\section{Capacity region analysis}\label{capacity}
In this section we consider i.i.d. Gaussian channels and noise vectors. We also assume $M\geq N_1\geq N_2\geq\ldots\geq N_K$ which results in $r_i = N_i (i\in[1:K])$ and therefore, $\tilde{\textbf{Y}}_i^n=\textbf{Y}_i^n$. From Fano's inequality,
\begin{align}
\sum_{i=1}^K\frac{nR_i}{r_i}  &\leq  \sum_{i=1}^K\frac{I(W_i;\textbf{Y}_i^n|\Omega^n,W_{[i+1:K]})}{r_i} \nonumber
\end{align}
\begin{align}
&\leq \frac{h(\textbf{Y}_K^n|\Omega^n)}{r_K}- \underbrace{\frac{h(\textbf{Y}_1^n|\Omega^n,W_{[1:K]})}{r_1}}_{n\log (2\pi e)}\\
&\ \ \ +\underbrace{\sum_{i=1}^{K-1}\!\left[\!\frac{h(\textbf{Y}_i^n|\Omega^n,W_{[i+1:K]})}{r_i}-\!\!\right.
\left. \frac{h(\textbf{Y}_{i+1}^n|\Omega^n,W_{[i+1:K]})}{r_{i+1}}\!\!\right]}_{\leq 0}\!\!\label{e29}
\end{align}
where the last non-positive term is a result of the lemma in the previous section. From the above results, we get an outer bound for the achievable rate region as
\begin{equation}\label{e27.5}
  \sum_{i=1}^K\frac{R_i}{r_i}\leq \frac{h(\textbf{Y}_K^n|\Omega^n)}{nr_K}-\log(2\pi e).
\end{equation}
Therefore, an outer bound for the ergodic capacity region is
\begin{equation}\label{e27.75}
  \sum_{i=1}^K\frac{R_i}{r_i}\leq \frac{\max_{\mathbf{\Sigma}_{X}:\mbox{tr}(\mathbf{\Sigma}_{X})\leq P}E\left[\log \det(\textbf{I}_{r_K}+\textbf{H}_K^H\mathbf{\Sigma}_X\textbf{H}_K)\right]}{r_K}
\end{equation}
and since the channels have i.i.d. Gaussian elements, the optimal input covariance matrix is $\frac{P}{M}\textbf{I}_M$ \cite{Telatar}. Hence,
\begin{align}\label{e28}
  C^o(P)=\{&(R_1,R_2,\ldots,R_K)\in R_{\geq 0}^K| \nonumber \\&R_i\leq E\left[\log \det(\textbf{I}_{r_i}+\frac{P}{M}\textbf{H}_i^H\textbf{H}_i)\right]\ \forall i\nonumber\\&\sum_{i=1}^K\frac{R_i}{r_i}\leq \frac{E\left[\log \det(\textbf{I}_{r_K}+\frac{P}{M}\textbf{H}_K^H\textbf{H}_K)\right]}{r_K}\}
\end{align}
It is obvious that the outer bound is more affected by the
capacity of the point-to-point link from the transmitter to the
user with the lowest number of receive antennas.

\textbf{Definition.} We define a class of channels (a set of matrices) $\Theta(p,q,m)$ where each channel (matrix) in this class has its elements drawn from the distribution $p$ in such a way that the optimal input covariance matrix for achieving the capacity of the point-to-point link from the transmitter to the virtual user defined by this channel is diagonal with equal entries. The details for this condition are given in \cite[Exercise 8.6]{Tse}. We also assume that for each channel in this class, all the singular values have the distribution $q$. In other words,
\begin{align}
  &\Theta(p,q,m)=\left\{H\in C^{m\times n}\ \forall n\leq m|\mbox{ Elements of }H\sim p,\nonumber\right.\\&\arg\max_{\mathbf{\Sigma}_{X}:\mbox{tr}(\mathbf{\Sigma}_{X})\leq P}E\left[\log \det(\textbf{I}_{n}+H^H\mathbf{\Sigma}_XH)\right]=\frac{P}{m}\textbf{I}_m , \nonumber\\&\left.\mbox{and }\lambda_i(H^HH)\sim q, \forall i=1,\ldots,\mbox{rank}(H)\right\}.
\end{align}
\begin{theorem}
In a $K$-user Gaussian MIMO BC with $M\geq N_1\geq N_2\geq\ldots\geq N_K$ and all the channels from the class of $\Theta(p,q,M)$, the capacity region with CDIT is given by
\begin{equation}\label{e30}
  C(P)=\left\{(R_{[1:K]})\in R_{\geq 0}^K|\sum_{i=1}^K\frac{R_i}{r_i}\leq E_q\left[\log (1+\frac{P}{M}\lambda)\right]\right\}
\end{equation}
where $E_q\left[\log (1+\frac{P}{M}\lambda)\right]=\int \log (1+\frac{P}{M}x)q(x)dx.$
\end{theorem}
\begin{IEEEproof} According to (\ref{e27.75}) and the properties of $\Theta(p,q,M)$, we have
\begin{equation}
\sum_{i=1}^K\frac{R_i}{r_i}\leq \frac{\sum_{i=1}^{r_K}E\left[\log (1+\frac{P}{M}\lambda_i(\textbf{H}_K^H\textbf{H}_K))\right]}{r_K}.
\end{equation}
If the singular values of $\textbf{H}_K$ have the same distribution, we can write
\begin{equation}
\sum_{i=1}^K\frac{R_i}{r_i}\leq E\left[\log (1+\frac{P}{M}\lambda_1(\textbf{H}_K^H\textbf{H}_K))\right].
\end{equation}
Also, if the singular values have the same distribution across the users, the outer bound is easily achieved by orthogonal transmission strategies, and therefore it is the optimal capacity region.
\end{IEEEproof}
A special case of theorem 2 was shown for the two user Gaussian MIMO BC in \cite{Huang1}, in which all the eigenvalues of $\textbf{H}_k^H\textbf{H}_k (k=1,2)$ are unity.
\section{mimo interference channel with no csit}\label{IC}
Consider a $K$-user MIMO IC with $K$ transmitters and $K$ receivers equipped with $M_i$ and $N_i$ antennas, respectively ($i=1,2,\ldots,K$). The input-output relationship at channel use $t$ is given by
\begin{equation}
  \textbf{{Y}}_i(t)=\sum_{j=1}^K\textbf{{H}}_{i,j}^H(t)\textbf{{X}}_j(t)+\textbf{{Z}}_i(t)\ ,i\in[1:K]\ ,t\in[1:n]
\end{equation}
where $\textbf{{Y}}_i(t)$ is the received signal at receiver $i$, $\textbf{{H}}_{i,j}$ is the channel matrix from the transmitter $j$ to the receiver $i$, $\textbf{{X}}_j(t)$ is the transmitted vector by the transmitter $j$ satisfying $\sum_{t=1}^{n}\|\mathbf{x}_j(t)\|^2\leq nP$ and $\textbf{{Z}}_i(t)$ is the noise vector at the receiver $i$. We assume that the channels are drawn from the same distribution, while the noise vectors could have different distributions. We also assume perfect CSIR (each receiver knows all the incoming channels to it from all the transmitters) and no CSIT.

For the two user case, theorems 2 and 3 in \cite{Huang1} are combined into theorem 5 in \cite{Vaze}. Here, we provide an alternative proof for it. We assume $N_1\leq N_2$ and $r_i=\min(M_2,N_i)$. By giving the message of user 1 to user 2, we have
\begin{align}
  \frac{nR_1}{r_1}+\frac{nR_2}{r_2} &\leq \frac{I(W_1;\tilde{\textbf{Y}}_1^n|\Omega^n,\Lambda^n)}{r_1}+\frac{I(W_2;\tilde{\textbf{Y}}_2^n|\Omega^n,\Lambda^n,W_{1})}{r_2}\nonumber  \\
  &= \frac{h(\tilde{\textbf{Y}}_1^n|\Omega^n,\Lambda^n)}{r_1}-o(\log P) \nonumber \\
  &\ \ \ +\frac{h(\tilde{\textbf{Y}}_2^n|\Omega^n,\Lambda^n,W_{1})}{r_2}-\frac{h(\tilde{\textbf{Y}}_1^n|\Omega^n,\Lambda^n,W_{1})}{r_1}
\end{align}
\begin{align}
  &\leq \frac{n\min(N_1,M_1+M_2)}{r_1}\log P \nonumber \\
  &\ \ \ +\underbrace{\frac{r_1h(\textbf{Y}_2^n|\Omega^n,\Lambda^n,W_{1})-r_2h(\textbf{Y}_1^n|\Omega^n,\Lambda^n,W_{1})}{r_1r_2}}_{\leq 0}\label{e42}
  %&\leq \frac{n\min(N_1,M_1+M_2)}{r_1}\log P
\end{align}
where $\tilde{\textbf{Y}}_1^n$ and $\tilde{\textbf{Y}}_2^n$ are the same as those in (\ref{e3}) and we have neglected all the terms with $o(\log P)$. In (\ref{e42}), $h(\tilde{\textbf{Y}}_1^n|\Omega^n,\Lambda^n)$ is maximized when $\tilde{\textbf{Y}}_1^n$ is Gaussian received from a transmitter with $M_1+M_2$ antennas. Also, in the term $[\frac{h(\tilde{\textbf{Y}}_2^n|\Omega^n,\Lambda^n,W_{1})}{r_2}-\frac{h(\tilde{\textbf{Y}}_1^n|\Omega^n,\Lambda^n,W_{1})}{r_1}]$, since the entropies are conditioned on $W_1$, the values of $\textbf{X}_1(1),\textbf{X}_1(2),\ldots,\textbf{X}_1(n)$ are known. Therefore, the extensions of $\textbf{H}_{11}^H(t)\textbf{X}_1(t)$ and $\textbf{H}_{21}^H(t)\textbf{X}_1(t)$ over $n$ channel uses can be removed from $\tilde{\textbf{Y}}_1^n$ and $\tilde{\textbf{Y}}_2^n$, respectively. What remains is a broadcast channel with a transmitter having $M_2$ transmit antennas. With a difference of $o(\log P)$, we can replace $\tilde{\textbf{Y}}_1^n$ and $\tilde{\textbf{Y}}_2^n$ with their linearly independent elements $\textbf{Y}_1^n$ and $\textbf{Y}_2^n$, respectively as in (\ref{e4}). Since $r_1\leq r_2$, following the same approach as in the formulae (\ref{e15.75}) to (\ref{e18}), we get the non-positive term in (\ref{e42}). Therefore, the outer bound is
\begin{align}
  D^o=\{&(d_1,d_2)\in R_{\geq 0}^2|\ d_i\leq \min(M_i,N_i)\ i=1,2\mbox{ and }\nonumber\\
  &\frac{d_1}{r_1}+\frac{d_2}{r_2}\leq\frac{\min(N_1,M_1+M_2)}{r_1}\}.
\end{align}
For the $K$-user IC, following the same proof in this paper for the broadcast channel, an outer bound for the DoF region can be obtained if the transmitters cooperate to make a broadcast channel (with $M_T=\sum_{i}M_i$). Therefore,
\begin{align}
  D^o=\{&(d_1,d_2,\ldots,d_K)\in R_{\geq 0}^K|\nonumber\\&d_i\leq \min(M_i,N_i)\ \forall i\mbox{ and } \sum_{i=1}^K\frac{d_i}{\min(M_T,N_i)}\leq 1\}.
\end{align}
According to theorem 9 in \cite{Vaze}, the above outer bound is tight provided that either $N_i\leq M_i\ \forall i$ or $N_i=N \geq M=M_i \ \forall i$ where in the former time sharing across the users and in the latter receive zero-forcing and time sharing are the achievable schemes, respectively.
\section{Conclusion}\label{conc}
In this paper, a new proof for the DoF region of the $K$-user MIMO BC with no CSIT was provided. The advantage of this proof is in finding the capacity region of a specific class of the $K$-user Gaussian MIMO BC with CDIT as explored in section \ref{capacity}.
Also, an outer bound for the DoF region of the $K$-user MIMO IC with no CSIT was provided.
\bibliography{REFERENCE1}
\bibliographystyle{IEEEtran}
\end{document}